\UseRawInputEncoding

\documentclass[journal]{IEEEtran}
\IEEEoverridecommandlockouts
\usepackage{amsmath,amssymb,amsfonts}
\usepackage{cite}
\usepackage{hyperref}
\usepackage{subcaption}
\usepackage{tikz}

\usepackage{helvet} 
\usepackage{sansmath} 
\usepackage{graphicx}
\usepackage{float}

\usepackage{titlesec}
\usetikzlibrary{positioning}

\title{Smell of Source:\  Learning-Based Odor Source Localization with Molecular Communication}

\author{
    Ayse Sila Okcu, \IEEEmembership{Student Member, IEEE}, and Ozgur B. Akan, \IEEEmembership{Fellow, IEEE} \\
    \thanks{Ayse Sila Okcu is with the Internet of Everything Group, Electrical Engineering
    Division, Department of Engineering, University of Cambridge, CB3 0FA
    Cambridge, U.K. (e-mail: aso32@cam.ac.uk).
    
    Ozg\"{u}r B. Akan is with the Internet of Everything Group, Electrical Engineering Division, Department of Engineering, University of Cambridge, CB3 0FA Cambridge, U.K., and also with the Center for neXt-Generation Communications (CXC), Department of Electrical and Electronics Engineering, Ko\c{c} University, 34450 Istanbul, Turkey (e-mail: oba21@cam.ac.uk).
}
}

\begin{document}
\maketitle
\begin{abstract}

Odor source localization is a fundamental challenge in molecular communication, environmental monitoring, disaster response, industrial safety, and robotics. In this study, we investigate three major approaches: Bayesian filtering, machine learning (ML) models, and physics-informed neural networks (PINNs) with the aim of odor source localization in a single-source, single-molecule case. By considering the source-sensor architecture as a transmitter-receiver model we explore source localization under the scope of molecular communication. Synthetic datasets are generated using a 2D advection-diffusion PDE solver to evaluate each method under varying conditions, including sensor noise and sparse measurements. Our experiments demonstrate that \textbf{Physics-Informed Neural Networks (PINNs)} achieve the lowest localization error of \(\mathbf{0.89 \times 10^{-6}}\) m, outperforming \textbf{machine learning (ML) inversion} (\(\mathbf{1.48 \times 10^{-6}}\) m) and \textbf{Kalman filtering} (\(\mathbf{1.62 \times 10^{-6}}\) m). The \textbf{reinforcement learning (RL)} approach, while achieving a localization error of \(\mathbf{3.01 \times 10^{-6}}\) m, offers an inference time of \(\mathbf{0.147}\) s, highlighting the trade-off between accuracy and computational efficiency among different methodologies.
\end{abstract}

\begin{IEEEkeywords}
Odor Source Localization, Bayesian Filtering, Machine Learning, Physics-Informed Neural Networks, Molecular Communication.
\end{IEEEkeywords}

\section{Introduction}
\IEEEPARstart{O}{dor} source localization (OSL) is a critical challenge in molecular and nanoscale communications, environmental monitoring, disaster response, industrial safety, and robotics \cite{aktas2024odor, akan2016fundamentals, lilienthal2004building}. As an inverse problem, OSL requires detecting a chemical signal in a turbulent medium and inferring its source location despite unpredictable dispersion effects.

Various strategies have been proposed to address this problem. In robotics, Gaussian plume models combined with Bayesian filtering have been used for robust odor tracking under noisy conditions \cite{vergassola2007infotaxis, hayes2002distributed, farrell2002filament, farrell2005chemical}. Reinforcement learning (RL) approaches \cite{vergassola2007infotaxis, li2024active} enable agents to navigate turbulent environments without relying on direct gradient measurements. More recently, deep learning techniques including conventional neural networks and physics-informed neural networks (PINNs) \cite{raissi2019physics, raissi2020hidden, pang2020npinns} have been applied to solve forward and inverse problems in fluid dynamics, leading rapid source localization in molecular communication systems \cite{nakano2013molecular, akan2016fundamentals}.

Although many methods have been developed for macroscale environments, the underlying physical and mathematical principles apply equally to nano- and microscale systems. In molecular communication, the transmitter–receiver architecture mirrors OSL, as nanoscale sensors detect molecular signals emitted from localized sources \cite{akyildiz2019information, nakano2013molecular}. This has direct applications in biomedical systems and nanonetworks, such as chemical sensing in microfluidic channels and in vivo tracking of signaling molecules.

In this work, we frame OSL within the molecular communication paradigm and compare various approaches: Bayesian and learning-based models. Synthetic datasets generated using a 2D advection–diffusion PDE solver \cite{smith1985numerical, farrell2002filament} are employed to evaluate each method under varying noise and sensor conditions. Our quantitative results highlight trade-offs in computational efficiency, accuracy, and robustness, thereby demonstrating the potential of these techniques for molecular-scale odor source localization.

\section{Methods}
In MC-based networks, nanomachines or microscale sensors attempt to locate a signaling molecule's origin, similar to how OSL techniques estimate an odor source’s position\cite{akan2016fundamentals,aktas2024odor}. To simulate these systems, we use the advection-diffusion equation, describing how molecules or odor particles propagate under diffusion and external forces (e.g., wind or fluid flow) as described in\cite{farrell2002filament,smith1985numerical}. 
Physical modeling using the advection-diffusion PDE to generate synthetic concentration fields similar to\cite{hayes2002distributed}.
As our main methods, we considered Bayesian methods to iteratively estimate the source location as described in\cite{vergassola2007infotaxis,linka2022bayesian} and Learning-based models to approximate odor concentration without solving PDEs using surrogate models as in\cite{pang2020npinns}.
All experiments were conducted in Python 3.8 with PyTorch for ML/PINNs applications, running on an Intel i7 CPU (16\,GB RAM) without GPU acceleration. The complete source code and data are publicly available\footnote{The source code and data can be accessed at: \url{https://github.com/aysesila/OSL-SingleSource}.}.

\subsection{Domain and Source Parameters} We consider a two-dimensional microfluidic channel of size $10\times10$\,\(\mu \text{m}^2\), discretized into a $50\times50$ grid. This environment simulates a  typical setting of lab-on-a-chip devices and microfluidic platforms used in biomedical applications. In such systems, chemical or molecular signals are often released from a localized transmitter (e.g., a nanomachine or a cell secreting signaling molecules) and detected by microscale sensors embedded within the channel. 

The diffusion coefficient is \(D=10^{-10}\,\text{m}^2/\text{s}\), a typical value for biological molecules \cite{akyildiz2019information}. A mild flow \(\mathbf{u}=(0.5,0)\,\mu\text{m}/\text{s}\) is used to include advection \cite{farrell2005chemical} (or \(\mathbf{u}=(0,0)\) for pure diffusion). We also add pulse injection and a first-order degradation term to capture realistic molecular communication dynamics. Although the model is two-dimensional, it effectively captures the key diffusion and advection dynamics observed in confined environments, such as microfluidic channels used for chemical sensing and biomedical assays.

\subsection{Mathematical Model of Odor Dispersion} To simulate odor or molecular dispersion, we model the concentration field using the 2D advection-diffusion PDE \cite{smith1985numerical}:
\begin{equation}\label{eq:PDE-1}
    \frac{\partial C}{\partial t} 
    = D \left( \frac{\partial^2 C}{\partial x^2} + \frac{\partial^2 C}{\partial y^2} \right)
    - u_x \frac{\partial C}{\partial x} 
    - u_y \frac{\partial C}{\partial y} 
    + S(x,y,t),
\end{equation}
where \(C(x,y,t)\) is the concentration at position \((x,y)\) and time \(t\); \(D\) is the diffusion coefficient; \((u_x,u_y)\) are the wind velocity components; and \(S(x,y,t)\) is the source term describing molecule/odor release at \((x_s,y_s)\):
\begin{equation}\label{eq:PDE-source}
    S(x,y,t) = Q \,\delta(x - x_s)\,\delta(y - y_s),
\end{equation}
where \(Q\) is the emission rate of molecules\cite{farrell2002filament}.
\begin{figure}[H]
    \centering
    \begin{subfigure}[b]{0.48\linewidth}
        \includegraphics[width=\textwidth]{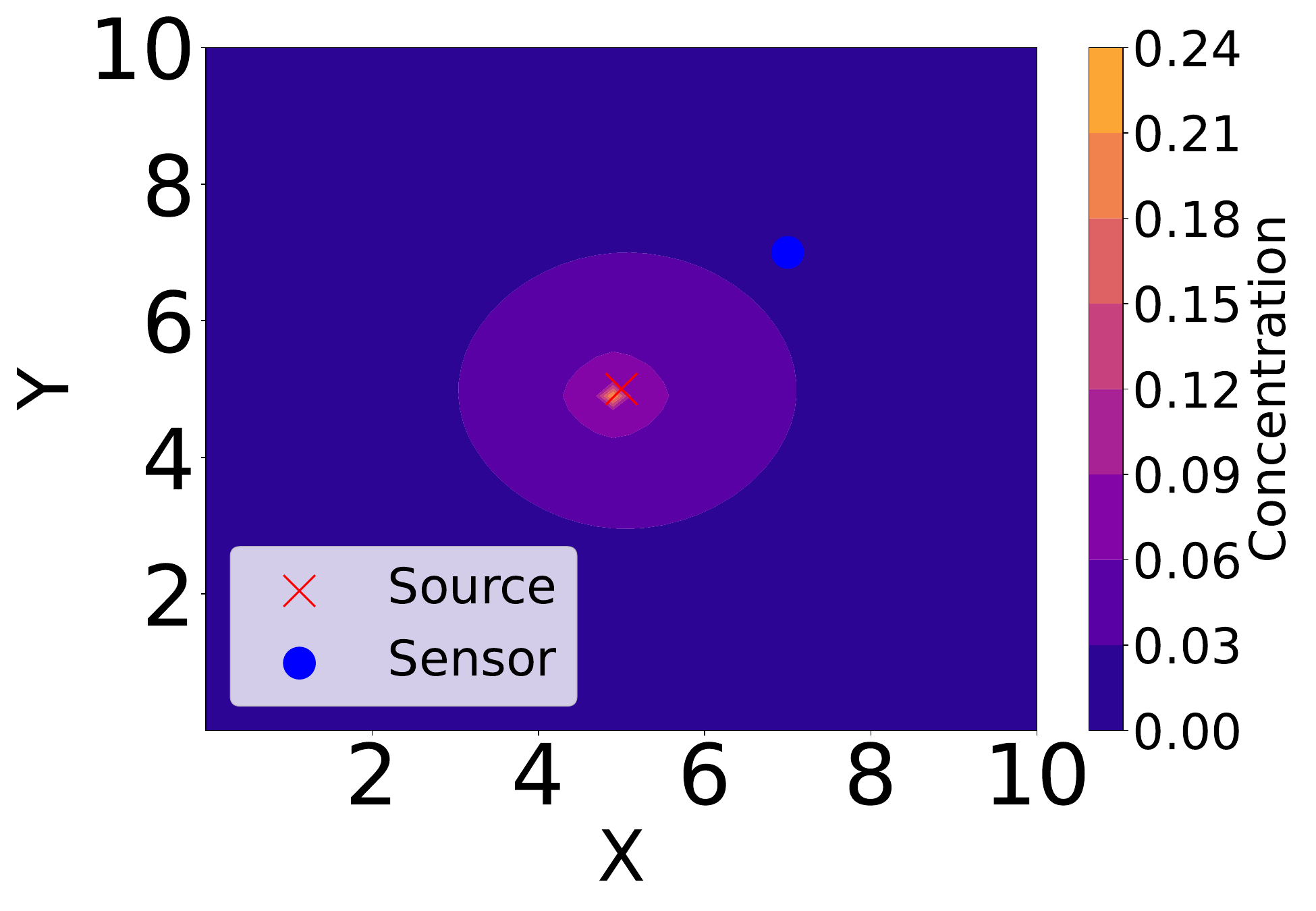}
        \caption{}
        \label{fig:conc1}
    \end{subfigure}
    \hspace{0.1cm}
    \begin{subfigure}[b]{0.48\linewidth}
        \includegraphics[width=\textwidth]{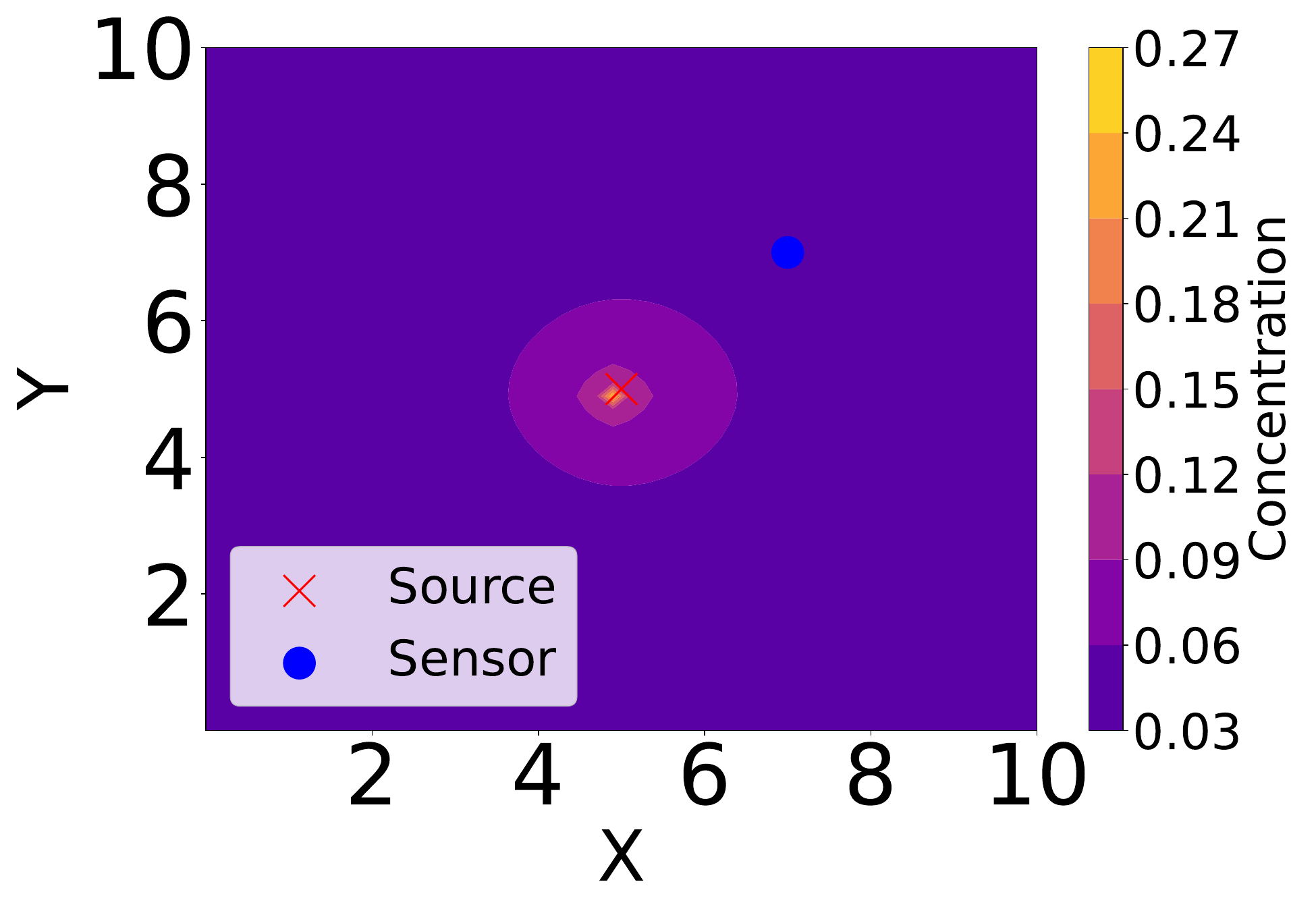}
        \caption{}
        \label{fig:conc2}
    \end{subfigure}
    \vspace{0.1cm}
    \begin{subfigure}[b]{0.48\linewidth}
        \includegraphics[width=\textwidth]{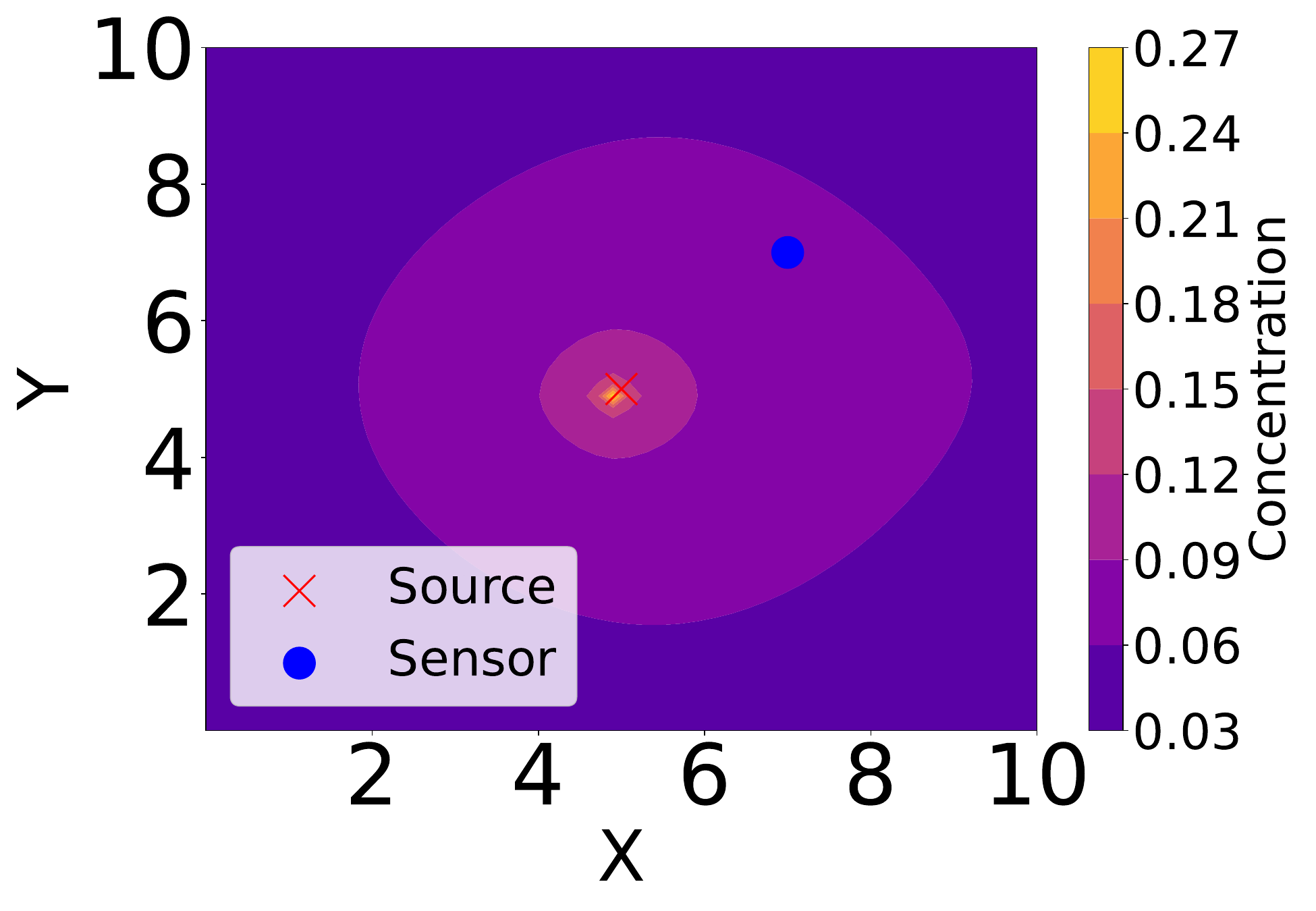}
        \caption{}
        \label{fig:conc3}
    \end{subfigure}
    \hspace{0.1cm}
    \begin{subfigure}[b]{0.48\linewidth}
        \includegraphics[width=\textwidth]{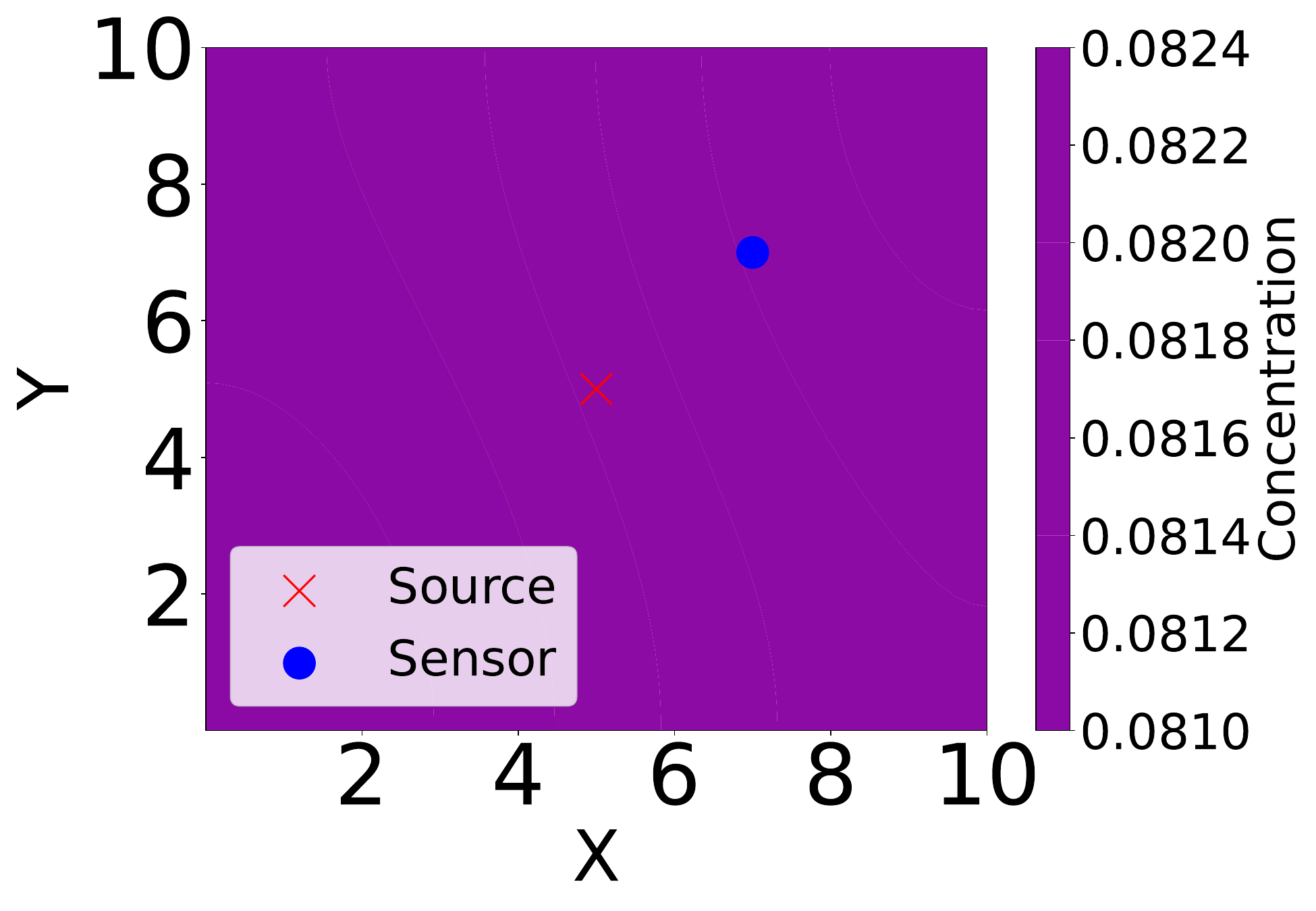}
        \caption{}
        \label{fig:conc4}
    \end{subfigure}
    \caption{Odor concentration field at different simulation times a) t=5s, b) t=10s, c) t=15s, d) t=27.5s . Higher concentrations are shown near the source location; advection moves the plume in the positive \(x\)-direction.}
    \label{fig:concentration_snapshots}
\end{figure}
In MC-based networks, nanomachines or microscale sensors attempt to locate a signaling molecule's origin, similar to how OSL techniques estimate an odor source’s position.

\subsection{Synthetic Data Generation}\label{sec:DataGeneration}
We solve a slightly modified advection-diffusion-reaction equation in time \cite{smith1985numerical, farrell2002filament}:
\begin{equation}
    \frac{\partial C}{\partial t} 
    + \mathbf{u} \cdot \nabla C
    = D\,\nabla^2 C 
    - \lambda C
    + S(x,y,t),
\end{equation}
where a pulse injection (active for \(t \leq T_{\text{inj}}\)) and a first-order degradation term \(-\lambda C\) are incorporated. Appropriate boundary conditions (Dirichlet: \(C=0\) or reflective/Neumann) are applied\cite{smith1985numerical}. The solver iterates with a small time step \(\Delta t\) until we reach a steady or stable solution. The final concentration field, a \(50\times 50\) matrix \(C(x,y)\), is optionally perturbed by additive Gaussian noise \(\epsilon \sim \mathcal{N}(0,\sigma^2)\), with \(\sigma\) set to 10\% of the maximum concentration to mimic sensor inaccuracies.

A subset of grid points---the location(s) of one or more sensors---is used to extract ``observed'' concentrations similar to those in \cite{lilienthal2004building}. For instance, if the simulation is configured with a sensor at \((5,5)\,\mu\text{m}\), the corresponding grid value in the solution matrix is the synthetic observation. This synthetic data serves as a common ground-truth for the Bayesian and ML-based methods.

\section{Implementation of Localization Methods}
\subsection{Bayesian Formulated Methods} Bayesian inference is a widely used approach for source localization, estimating the unknown source location \(\mathbf{x}_s = (x_s,y_s)\) given a set of sensor measurements \(\mathbf{z} = \{z_1,\ldots,z_T\}\). 

By Bayes' theorem, the posterior probability of the source location given sensor measurements is:
\begin{equation}
    p(\mathbf{x}_s \mid \mathbf{z}) \propto p(\mathbf{z} \mid \mathbf{x}_s) p(\mathbf{x}_s),
\end{equation}
where:
\begin{itemize}
  \item \(p(\mathbf{x}_s)\) is the \textbf{prior} distribution representing initial beliefs about the source location.
  \item \(p(\mathbf{z} \mid \mathbf{x}_s)\) is the \textbf{likelihood} function, which models the probability of obtaining sensor readings given a candidate source.
\end{itemize}
\noindent
\textbf{Prior Distribution:} \noindent In the absence of prior information, we assume a uniform prior over the search domain. 
\newline
\textbf{Likelihood Function:} \noindent To model sensor noise, we assume that the observed odor concentration follows a Gaussian distribution around the expected concentration from the odor dispersion model:
\begin{equation}
    p(z_i \mid \mathbf{x}_s) = 
    \mathcal{N}\bigl(z_i \mid c(\mathbf{x}_i;\mathbf{x}_s),\,\sigma^2 \bigr).
\end{equation}
\noindent
where \(z_i\) is the measured concentration at sensor location \(\mathbf{x}_i\), 
\(c(\mathbf{x}_i; \mathbf{x}_s)\) is the expected concentration given a source at \(\mathbf{x}_s\), 
and \(\sigma^2\) is the noise variance. 
When sensor data are collected over multiple time steps \(t=1,\ldots,T\), the likelihood extends to
\[
    p(\mathbf{z}_{1:T} \mid \mathbf{x}_s) 
    \;=\; \prod_{t=1}^{T} \mathcal{N}\bigl(z_{t} \mid c(\mathbf{x}_t;\mathbf{x}_s),\sigma^2 \bigr).
\]

\subsubsection{ Maximum A Posteriori (MAP) Estimation} 
In the classical Bayesian approach, we compute the posterior on a discretized grid and pick the location with maximum posterior probability:
\begin{equation}
    \mathbf{x}_s^* = \arg\max_{\mathbf{x}_s} p(\mathbf{x}_s \mid \mathbf{z}).
\end{equation}
For a uniform prior, the posterior simplifies to:
\begin{equation}
    p(\mathbf{x}_s \mid \mathbf{z}) \propto p(\mathbf{z}\mid \mathbf{x}_s).
\end{equation}
\noindent
A grid search over all candidate locations identifies the source estimate. This method is robust to noise but computationally demanding for high-resolution grids \cite{vergassola2007infotaxis}.

\begin{figure}[H]
    \centering
    \begin{subfigure}[b]{0.48\linewidth}  
        \centering
        \includegraphics[width=\textwidth]{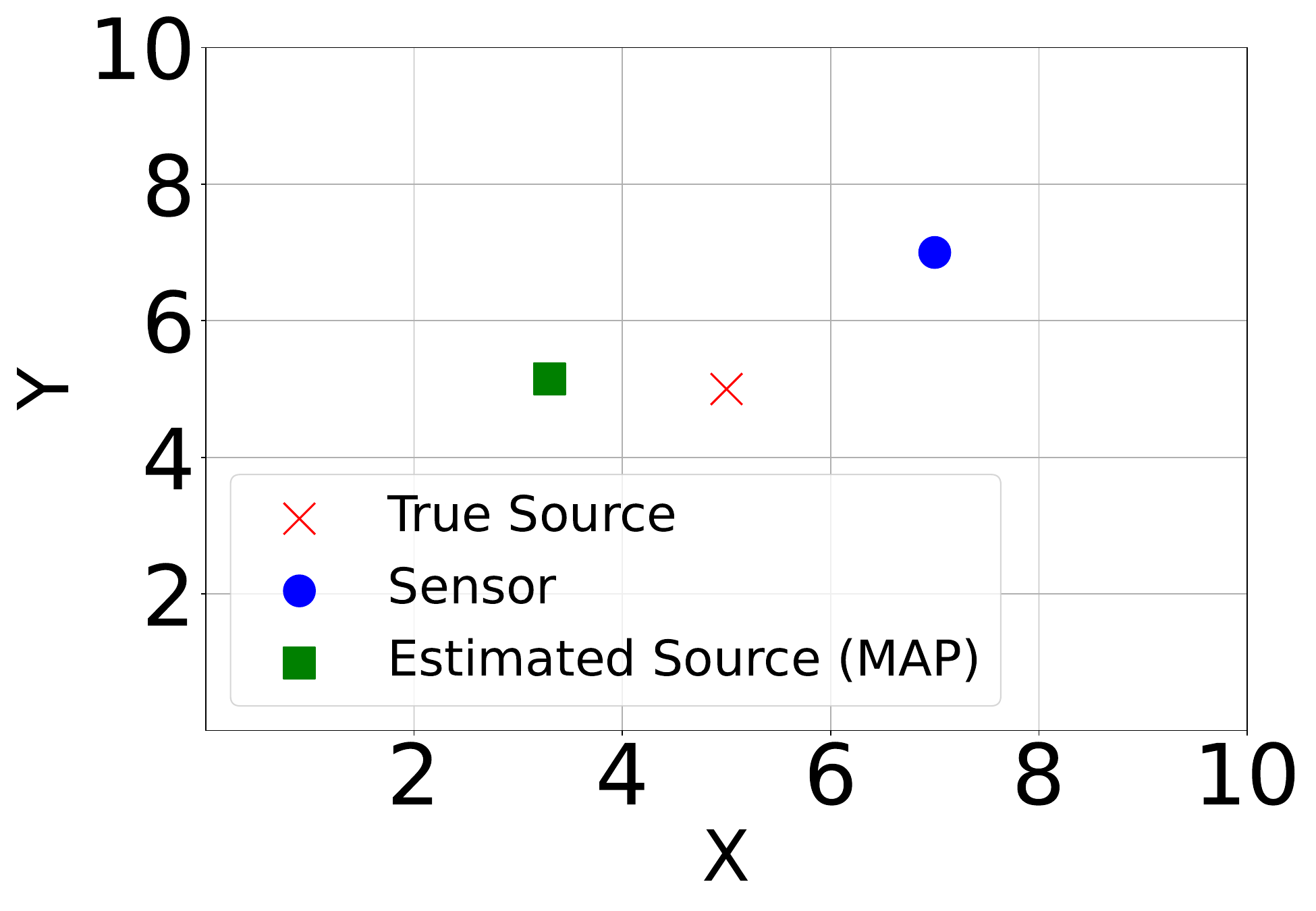}
        \caption{}
        \label{fig:bayes1}
    \end{subfigure}
    \hspace{0.1cm} 
    \begin{subfigure}[b]{0.48\linewidth}  
        \centering
        \includegraphics[width=\textwidth]{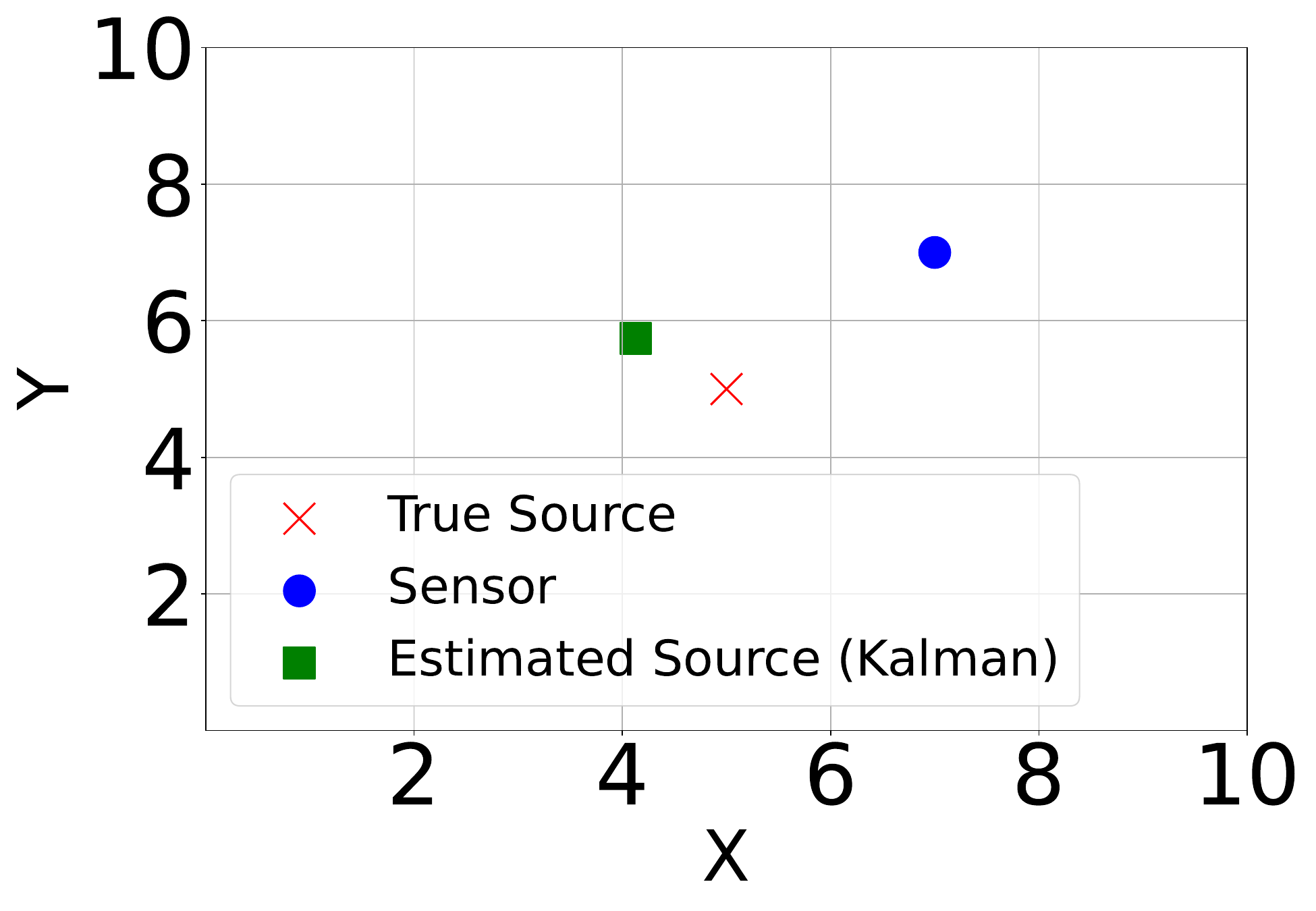}
        \caption{}
        \label{fig:bayes2}
    \end{subfigure}
    \caption{Bayesian estimation results: (a) MAP on a grid, (b) Kalman filtering over time.}
    \label{fig:bayes}
\end{figure}
\subsubsection{Kalman Filter (KP)}
While MAP estimation yields a single-shot estimate, Bayesian filtering methods \emph{sequentially} 
update the posterior as new sensor readings arrive, suitable for dynamic or real-time odor localization.
When the system dynamics are linear and noise is Gaussian, a Kalman Filter can be applied. Here, the state \(\mathbf{x}_k=[x_s,y_s]^T\) remains stationary (static source), and the measurement model relates sensor concentration to the PDE-based concentration fields \cite{farrell2005chemical}.

\subsection{Learning-Based Methods} 
Our study investigates: 
(1) a multilayer perceptron (MLP), (2) physics-informed neural networks (PINNs) and (3) reinforcement learning (RL).


\subsubsection{Multi Layer Perceptron (MLP)}
Instead of explicitly inverting the forward model of odor dispersion, the MLP is trained to learn a direct mapping from sensor measurements and wind parameters to the \emph{relative source offset}. Given a (down-sampled) time series of sensor concentration data \(\mathbf{z}\) and the wind vector \(\mathbf{u}\), the network outputs an estimate \(\Delta \mathbf{x}\) that represents the displacement from the sensor's location to the odor source. Hence, the \emph{absolute} source location is computed as:
\[
\hat{\mathbf{x}}_s = \mathbf{x}_{\text{sensor}} + \Delta \mathbf{x}.
\]
\noindent
Let \(f_\theta(\cdot)\) denote the neural network. We construct a training set where each sample consists of sensor readings (with noise) and the corresponding true relative offset, 
\[
\Delta \mathbf{x}^{\mathrm{true}} = \mathbf{x}_s^{\mathrm{true}} - \mathbf{x}_{\text{sensor}}.
\]
The loss function is defined as the mean squared error (MSE):
\[
\min_{\theta} \sum_{i=1}^{N} \left\| f_\theta(\mathbf{z}_i, \mathbf{u}_i) - \Delta \mathbf{x}_i^{\mathrm{true}} \right\|^2.
\]
We train the model on 4,000 samples (80\% training, 20\% testing) using the Adam optimizer with a learning rate of \(10^{-3}\).

The MLP architecture is defined as follows:
\begin{itemize}
    \item \textbf{Input Layer}: 602 inputs (600 sensor readings + 2 wind components).
    \item \textbf{Hidden Layers}: Three fully connected layers with 256, 128, and 64 neurons respectively, each employing ReLU activations.
    \item \textbf{Output Layer}: A 2-dimensional output representing the relative offset \(\Delta \mathbf{x}\).
\end{itemize}
At inference, the network predicts the relative offset, which is then added to the sensor’s location to yield the estimated source location. 

\subsubsection{Physics-Informed Neural Networks (PINNs)} 
PINNs embed the underlying physics of odor dispersion modeled by the advection-diffusion PDE—directly into the training objective. In our formulation, the network \(c_\theta(x,y)\) is used to approximate the concentration field, and the source location \(\mathbf{x}_s\) is treated as an additional learnable parameter. By incorporating a physics loss that penalizes the PDE residual, the network is guided to produce outputs that are both data-consistent and physically plausible.
 \noindent
The total loss is defined as:
\[
L_{\text{total}}(\theta, \mathbf{x}_s) = \lambda_{\mathrm{MSE}} \, L_{\mathrm{MSE}} + \lambda_{\mathrm{phy}} \, L_{\mathrm{physics}},
\]
where
\[
L_{\mathrm{MSE}} = \sum_{i=1}^{N} \left\| f_\theta(\mathbf{z}_i, \mathbf{u}_i) - \Delta \mathbf{x}_i^{\mathrm{true}} \right\|^2,
\]
and the physics loss is given by
\[
L_{\mathrm{physics}} = \left\| \nabla c_\theta(x,y) - \left(D\,\nabla^2 c_\theta(x,y) - \mathbf{u}\cdot\nabla c_\theta(x,y)\right) \right\|^2.
\]
Here, \(\lambda_{\mathrm{MSE}}\) and \(\lambda_{\mathrm{phy}}\) are weight factors for the data loss and the physics loss, respectively. We employ a 4-layer fully connected network. Automatic differentiation is used to compute the PDE residual, and both \(\theta\) and the source location \(\mathbf{x}_s\) are updated during training.

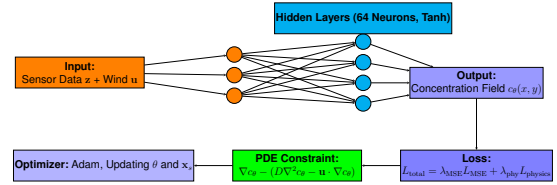
\begin{figure}
    \centering
    \resizebox{0.4\textwidth}{!}{ 
    \begin{tikzpicture}[
        node distance=15mm,
        every node/.style={draw, circle, minimum size=6mm, inner sep=0.5mm, font=\large\sffamily},
        every path/.style={->,>=stealth},
        box/.style={draw, rectangle, minimum width=50mm, minimum height=12mm, align=center, font=\large\sffamily}
    ]
        \node[box, fill=orange] (input) at (-7,0) {\textbf{Input:} \\ Sensor Data $\mathbf{z}$ + Wind $\mathbf{u}$};
        
        \foreach \i in {1,2,3} {
            \node[fill=orange] (i\i) at (-1,1.5-\i*0.8) {};
            \draw[->] (input) -- (i\i);
        }

        \foreach \i in {1,2,3,4} {
            \node[fill=cyan] (h\i) at (4,2-\i*0.8) {};
        }
        \node[box, fill=cyan, above=3mm] at (4,1.2) {\textbf{Hidden Layers (64 Neurons, Tanh)}};

        \node[box, fill=blue!40, right=of h3] (output) {\textbf{Output:} \\ Concentration Field $c_\theta(x,y)$};

        \node[box, fill=blue!50, below=of output, yshift=-0.5cm] (loss) {\textbf{Loss:} \\ $L_{\mathrm{total}} = \lambda_{\mathrm{MSE}} L_{\mathrm{MSE}} + \lambda_{\mathrm{phy}} L_{\mathrm{physics}}$};

        \node[box, fill=green, left=of loss] (pde) {\textbf{PDE Constraint:} \\ $\nabla c_\theta - (D \nabla^2 c_\theta - \mathbf{u} \cdot \nabla c_\theta)$};

        \node[box, fill=blue!30, left=of pde] (optimizer) {\textbf{Optimizer:} Adam, Updating $\theta$ and $\mathbf{x}_s$};

        \foreach \i in {1,2,3} {
            \foreach \h in {1,2,3,4} {
                \draw[->] (i\i) -- (h\h);
            }
        }
        \foreach \h in {1,2,3,4} {
            \draw[->] (h\h) -- (output);
        }
        \draw[->] (output) -- (loss); 
        \draw[->] (loss) -- (pde);
        \draw[->] (pde) -- (optimizer);

    \end{tikzpicture}
    }
    \caption{PINN Architecture}
    \label{fig:pinn_nn}
\end{figure}

The training data for the PINN are generated similarly to the MLP method, but the PINN loss also includes the physics term. Figure~\ref{fig:pinn_nn} outlines the PINN architecture, highlighting that the network receives sensor data, wind information, and implicitly the unknown source location, and outputs the relative source offset while enforcing both data fidelity and the PDE constraint.

\begin{figure}[htbp]
    \centering
    \begin{subfigure}[b]{0.48\linewidth}
        \centering
        \includegraphics[width=\textwidth]{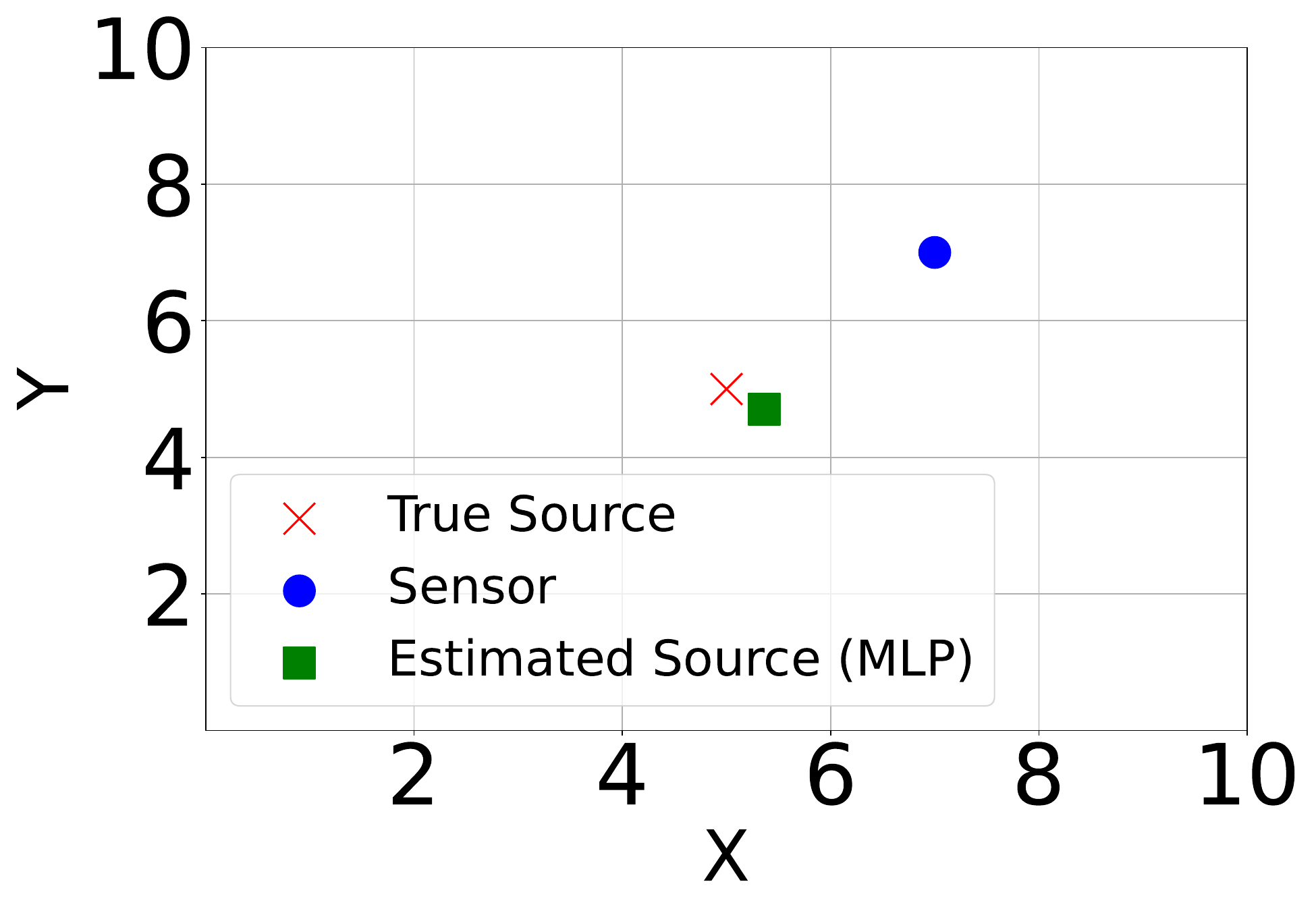}
        \caption{}
        \label{fig:mlp}
    \end{subfigure}
    \hspace{0.1cm}
    \begin{subfigure}[b]{0.48\linewidth}
        \centering
        \includegraphics[width=\textwidth]{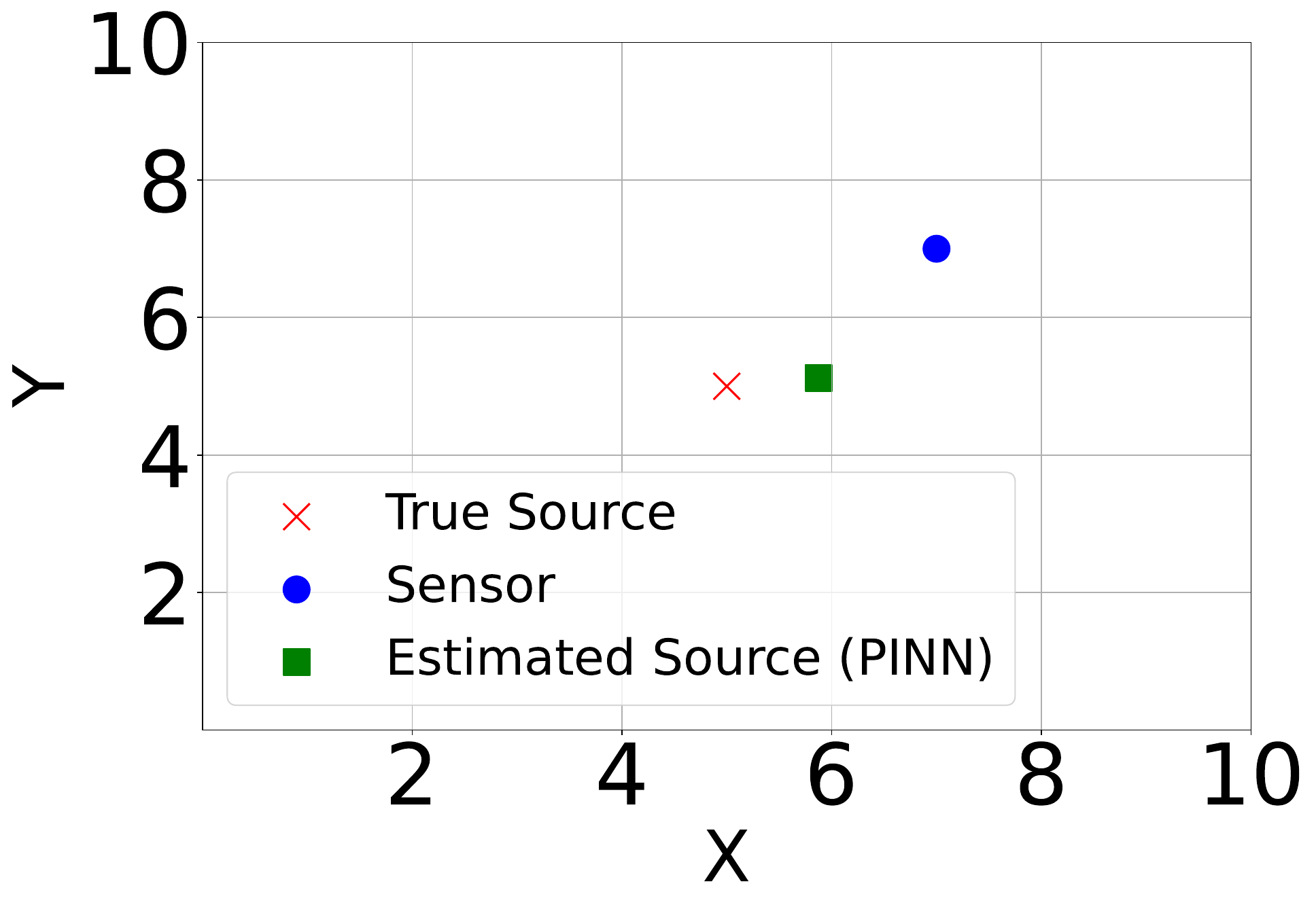}
        \caption{}
        \label{fig:pinns}
    \end{subfigure}
    \caption{Estimate results for a) MLP and b) PINNs estimates}
    \label{fig:learning}
\end{figure}

\subsubsection{Reinforcement Learning (RL)}
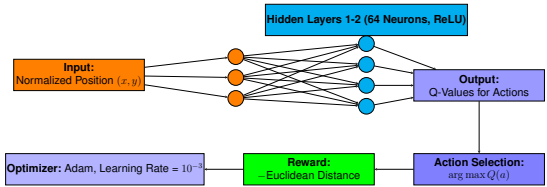
\begin{figure}[H]
    \centering
    \resizebox{0.4\textwidth}{!}{ 
    \begin{tikzpicture}[
        node distance=15mm,
        every node/.style={draw, circle, minimum size=6mm, inner sep=0.5mm, font=\large\sffamily}, 
        every path/.style={->,>=stealth},
        box/.style={draw, rectangle, minimum width=50mm, minimum height=12mm, align=center, font=\large\sffamily} 
    ]
        \node[box, fill=orange] (input) at (-7,0) {\textbf{Input:} \\ Normalized Position $(x, y)$};

        \foreach \i in {1,2,3} {
            \node[fill=orange] (i\i) at (-1,1.5-\i*0.8) {};
            \draw[->] (input) -- (i\i);
        }

        \foreach \i in {1,2,3,4} {
            \node[fill=cyan] (h\i) at (4,2-\i*0.8) {};
        }
        \node[box, fill=cyan, above=3mm] at (4,1.2) {\textbf{Hidden Layers 1-2 (64 Neurons, ReLU)}};

        \node[box, fill=blue!40, right=of h3] (output) {\textbf{Output:} \\ Q-Values for Actions};

        \node[box, fill=blue!50, below=of output, yshift=-0.5cm] (action) {\textbf{Action Selection:} \\ $\arg\max Q(a)$};

        \node[box, fill=green, left=of action] (reward) {\textbf{Reward:} \\ $-\text{Euclidean Distance}$};

        \node[box, fill=blue!30, left=of reward] (optimizer) {\textbf{Optimizer:} Adam, Learning Rate = $10^{-3}$};

        \foreach \i in {1,2,3} {
            \foreach \h in {1,2,3,4} {
                \draw[->] (i\i) -- (h\h);
            }
        }
        \foreach \h in {1,2,3,4} {
            \draw[->] (h\h) -- (output);
        }
        \draw[->] (output) -- (action); 
        \draw[->] (action) -- (reward);
        \draw[->] (reward) -- (optimizer);

    \end{tikzpicture}
    }
    \caption{DQN Architecture}
    \label{fig:rl_architecture}
\end{figure}

To further explore active localization strategies, we implemented a reinforcement learning (RL) approach based on a Deep Q-Network (DQN).  In our RL framework, the environment is modeled as a discrete 2D grid (representing a microfluidic channel with a fixed odor source. The agent’s state is its normalized grid position, and it selects one of four actions (up, down, left, right) to reduce its Euclidean distance to the source. Our DQN consists of two hidden layers (64 neurons each, ReLU activations) and an output layer that yields Q-values for the actions (see Figure~\ref{fig:rl_architecture}).

\
During training, an epsilon-greedy policy with experience replay is employed. The training loop runs for 500 episodes, and at each step the agent’s state is updated, the transition stored in a replay buffer, and a mini-batch is sampled to update the network using the Adam optimizer. Finally, we convert the agent’s final grid position into normalized coordinates (by dividing by the grid size) to compute the localization error against the known true source location.

\section{Experimental Results and Discussion}
Table~\ref{tab:localization} summarizes localization performance for various approaches. We highlight key findings from each method.

\begin{table}[htbp]
    \centering
    \caption{Localization Performance Comparison}
    \begin{tabular}{lccc}
        \hline
        \textbf{Method} & \textbf{Estimated Source} & \textbf{True Source} & \textbf{Error (m)} \\
        \hline
        MAP & [3.55, 5.27] \(\times 10^{-6}\) & [5.0, 5.0] \(\times 10^{-6}\) & 1.48 \(\times 10^{-6}\) \\
        Kalman Filter & [4.17, 5.78] \(\times 10^{-6}\) & [5.0, 5.0] \(\times 10^{-6}\) & 1.62 \(\times 10^{-6}\) \\
        PINNs & [5.87, 5.11] \(\times 10^{-6}\) & [5.0, 5.0] \(\times 10^{-6}\) & 0.89 \(\times 10^{-6}\) \\
        RL & [6.0, 6.8] \(\times 10^{-6}\) & [3.0, 7.0] \(\times 10^{-6}\) & 3.01 \(\times 10^{-6}\) \\
        \hline
    \end{tabular}
    \label{tab:localization}
\end{table}

Overall, our experiments reveal key insights into odor source localization. Scale and normalization are critical, especially in ML and MAP estimation, which performed well but required careful handling at smaller scales. ML-based methods faced optimization challenges, including flat loss surfaces and gradient issues, where strategies like multi-start optimization could help. PINNs achieved the lowest error by incorporating the governing PDE, but balancing PDE regularization with data loss was essential. Bayesian filtering, such as the Kalman Filter, effectively processed sequential data but struggled with strong nonlinearities and sparse sensor coverage. Overall, while Bayesian inference and PINNs offer higher accuracy and better physical consistency, they are computationally intensive. In contrast, ML and RL methods enable faster real-time estimation but need improved optimization and normalization to match the performance of physics-driven approaches.

\section{Conclusion}
In this study, we compared odor source localization methods using a unified PDE-simulated dataset \cite{smith1985numerical, farrell2002filament}. We evaluated Bayesian (MAP, Kalman), ML (MLP inversion), PINNs, and RL approaches, highlighting trade-offs in accuracy, robustness, and computational overhead. Our findings suggest that PINNs improve physical consistency by enforcing PDE constraints, Bayesian methods provide robust estimates but can be computationally demanding while ML and RL methods offering fast inference but require careful input scaling and tuning. Future work will explore novel hybrid methods that combine physics-based models with data-driven techniques, as well as advanced RL strategies for active sensing in dynamic environments \cite{li2024active}. These results provide a roadmap for designing effective odor source localization strategies across multiple scales in molecular communication scenarios.


\end{document}